\documentclass{epl}
\title{ Mirages and enhanced magnetic interactions in quantum corrals}
\shorttitle{Mirages in quantum corrals}
\author{A. Correa\inst{1} \and K. Hallberg\inst{1} \and C. A. Balseiro\inst{1}}
\institute{
  \inst{1} Centro At\'omico Bariloche
and Instituto Balseiro, Comisi\'on Nacional de Energ\'{\i}a At\'omica, 8400 San
Carlos de Bariloche, Argentina.\\
}

\begin{document}

\maketitle

\begin{abstract}
 We develop a theory for the interactions between magnetic impurities in nanoscopic
 systems. The case of impurities in quantum corrals built on the (111) Cu surface is
  analyzed in detail. For elliptical corrals with one impurity, clear magnetic mirages are
 obtained. This leads to an enhancement of the inter-impurity interactions when two
impurities are placed at special points in the corral. We discuss the enhancement of the
 conduction electron response to the local perturbation in other nanoscopic systems.
\end{abstract}

\smallskip During the last decade, the physics of nanoscopic and mesoscopic
systems has generated an increasing interest in the community due to
technological advances that allow for controlled experiments and the
possible developments of new applications, nanoelectronics being one example%
\cite{gen}.

Coulomb blockade in small metallic or superconducting islands and in quantum
dots, the similarities between the physics of quantum dots and
Kondo impurities \cite{QD} or the observation of spectroscopic mirages in
quantum corrals with Kondo impurities \cite{Mano} make evident that in many
aspects interactions and correlations play a central role in the behavior of
these small systems. There are some ingredients that distinguish the physics
of nanoscopic systems from that of macroscopic samples. One of them, and the
most important in the context of the present work, is the quantification of
one electron levels due to confinement effects. Here we discuss the
interaction between magnetic impurities mediated by conduction electrons,
known as the RKKY interaction\cite{RKKY}. We analyze this mechanism in
nanoscopic systems and show that in quantum corrals there is a huge
enhancement of the impurity-impurity interaction due to confinement effects.

\begin{figure}
\onefigure{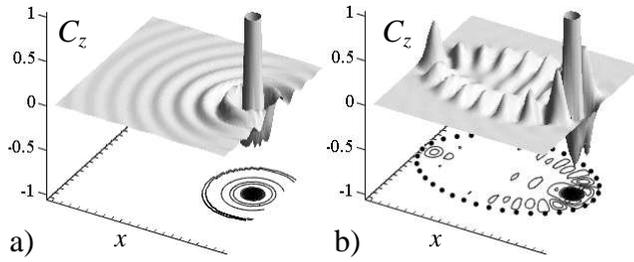}
\caption{Spin correlation functions $C_z$ between the impurity and
conduction electron 
spins (arbitrary units): a) open surface; b) elliptical quantum corral with 
the impurity at the right focus.} 
\label{fig1}
\end{figure}

Quantum corrals are built by positioning atoms, typically transition metal
atoms, along a closed line on the clean surfaces of noble metals. In recent
experiments Manoharan et. al.\cite{Mano} have built elliptical corrals with
Co atoms on the (111) surface of Cu. The Cu (111) surface has a band of
surface states, orthogonal to the bulk states, which can be represented as a
two dimensional electron gas confined at the surface. The Fermi level is
placed at 450 meV above the bottom of the surface state band. The atoms
forming the corral act as scattering centers which tend to confine surface
electrons inside the corral. If the corral fence where an impenetrable wall
electrons inside the corral would be perfectly confined and the energy
spectrum would consist of a set of delta functions at the energies of the
bound states. The characteristic energy separation between these states
decreases as the size of the corral increases. In real systems, there is some%
{\it \ leaking }of the wave function: electrons can tunnel through the 
fence and the bound states acquire a finite life time. The energy spectrum
inside the corral (the local density of states) then consists of
resonances, the width of which increases with increasing energy.
This structure can drastically change the response of the conduction
electrons to a local perturbation. Experiments with different transition metal ipurities 
inside corrals can be made. Among them, only Co impurities clearly show spectroscopic evidence of Kondo 
effect. Here we analyze the case of Fe or Ni impurities and describe them with an {\it s-d} Hamiltonian.
We show that the magnetic response is much more pronounced when the impurities are inside the corral 
rather than on an open surface and in some cases a magnetic mirage is
observed. To illustrate these effects we anticipate the results in fig.~\ref{fig1}
with the response of the two-dimensional (2D) surface band to a magnetic
impurity.
When the impurity, interacting via an exchange coupling with the
surface electrons, is at a free surface (fig. (1a)), the usual RKKY
oscillations in the correlations between the impurity and conduction
electron spins are observed. For the 2D case these oscillations decay as the
inverse square of the distance to the impurity. When the impurity is placed
at the focus of an elliptical corral with a principal axis of $180\AA $ and
eccentricity $\epsilon =0.75$, the polarization of the surface states is
much larger than for the previous case (fig. (1b), note that the scale in
figures (1a) and (1b) is the same). Moreover, close to the empty focus of
the elliptical corral there is a substantial maximum which could be
interpreted as the magnetic mirage of the impurity. These results indicate
than under some circumstances two magnetic impurities located at special
points of the corral could strongly interact despite of the fact that they
could be up to $100\AA $ apart. The effect is so strong that the
impurity-impurity interaction at these large distances could be even larger
that the interaction of two impurities placed at short distances on
an open surface.

These results where obtained by considering the following Hamiltonian:

\begin{equation}
H=H_{corral}+H_{imp}  \label{ham}
\end{equation}
where $H_{corral}$ describes the surface electrons in the presence of the
corral and is given by

\begin{equation}
H_{corral}=\sum_{\sigma }\int d{\bf r}\psi _{\sigma }^{\dagger }({\bf r})(-%
\frac{\hbar ^{2}}{2m^{*}}{\bf \nabla }^{2}+V_{corral}({\bf r}))\psi
_{\sigma }({\bf r})  \label{hcorr}
\end{equation}
here $\psi _{\sigma }^{\dagger }({\bf r})$ creates a conduction electron
with spin $\sigma $ at coordinate ${\bf r}$, $m^{*}=0.3m_{e}$ is the
effective mass of the surface electrons and $V_{corral}({\bf r})=\sum_{i}V(%
{\bf r}-{\bf R}_{i})$ is the potential generated by the atoms forming the
corral that are placed at coordinates ${\bf R}_{i}$. We consider typically
36 atoms in the corral fence as in the experiments of reference (3). The
second term in Hamiltonian (1) represents a magnetic impurity at coordinate $%
{\bf R}_{imp}$ that interacts with the conduction electron via an exchange
interaction $J$, and in usual notation it reads:

\begin{equation}
H_{imp}=-J{\bf S}_{imp}\cdot \psi ^{\dagger }({\bf R}_{imp}){\bf \hat{\sigma%
}}\psi ({\bf R}_{imp})  \label{himp}
\end{equation}
the operator ${\bf S}_{imp}$ describes the impurity spin, the components of%
{\bf \ }${\bf \hat{\sigma}}${\bf \ }are the Pauli matrices and $\psi
^{\dagger }({\bf r})=(\psi _{\uparrow }^{\dagger }({\bf r}),\psi
_{\downarrow }^{\dagger }({\bf r}))$. For the $sd$-model considered here,
the exchange coupling constant $J$ is positive ($J\simeq 1meV$).

We study diferent geometries: elliptical corrals with eccentricities $%
\epsilon =0$ , $0.5$ and $0.75$, and a square corral. To do so, we first
calculate the surface-electron propagator in the presence of the corral $%
G_{\sigma }^{0}({\bf r},{\bf r}^{\prime },\omega )$ that can be expressed in
terms of the free-electron propagator $g_{\sigma }({\bf r},{\bf r}^{\prime
},\omega )$ which in 2D is well known\cite{2Dg}; in the evaluation of this
quantity we have used a high energy cutoff. The results inside the ellipse
are not very sensitive to the shape of the scattering potential $V({\bf r})$ but
to the size and shape of the corral, that is to the coordinates ${\bf R}_{i}$
of the scattering centers. The results presented in this work were obtained
with $V({\bf r})=\alpha \delta ({\bf r})$ where $\delta ({\bf r})$ is the
Dirac delta function. The electron propagator is given by\cite{g0}:

\begin{equation}
G_{\sigma }^{0}({\bf r},{\bf r}^{\prime },\omega ) =g_{\sigma }({\bf r},%
{\bf r}^{\prime },\omega )+  
\alpha \sum_{ij}g_{\sigma }({\bf r},{\bf R}_{i},\omega )[{\bf 1}-\alpha 
{\bf g}]_{ij}^{-1}g_{\sigma }({\bf R}_{j},{\bf r}^{\prime },\omega )
\label{g0}
\end{equation}
here the matrix ${\bf g}$ is defined as ${\bf g}_{ij}=g_{\sigma }({\bf R}%
_{i},{\bf R}_{j},\omega )$.

The total density of states defined as

\begin{equation}
\rho ^{0}(\omega )=-\frac{1}{\pi }\sum_{\sigma }\int_{corral}d{\bf r}%
\mathop{\rm Im}%
G_{\sigma }^{0}({\bf r},{\bf r},\omega +i0^{+})  \label{tot}
\end{equation}
is shown in fig.(2a) and compared with the eigenstates of an isolated ellipse
with the same shape and dimensions\cite{Alig}. For low energies the
resonances obtained for the corral are narrow and coincide with the
eigenvalues of the isolated system. As the energy increases the resonances
become broader and are shifted. This is because for higher energies the
fence becomes more transparent, when the electron wave length $\lambda $ is
smaller than the distance $d$ between scattering centers, electrons can
easily tunnel through the fence. Figure (2a) also shows the density of
states for a corral of the same dimensions made with 24 scattering centers,
that has much broader resonances and at the Fermi energy the structure is
almost completely washed out. For the elliptical corral used in the rest of
the paper with $\epsilon=0.75$, which reproduces one of the experimental setups of Ref.
(3), the
Fermi level lies at one well defined resonance. As we discuss below, this
coincidence is important and determines the behavior of the system. Finally,
the local density of states at the Fermi energy $\rho ^{0}({\bf r}%
,\varepsilon _{F})=-\frac{1}{\pi }\sum_{\sigma }%
\mathop{\rm Im}%
G_{\sigma }^{0}({\bf r},{\bf r},\omega +i0^{+})|_{\omega =\varepsilon _{F}}$
is shown in figs. (2b) and (2c).

\begin{figure}
\onefigure{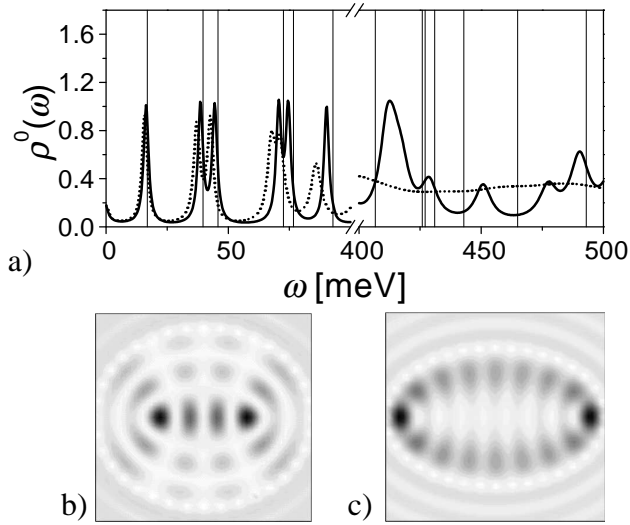}
\caption{a) Total density of states for the elliptical corral with
$\epsilon=0.5$ bounded by 36
(full line) and 24 atoms (dotted line). The vertical lines
correspond to the energies of a hard wall corral. b) and c): top view
of the local densities of states at the Fermi energy (450 meV) for elliptical 
corrals with 36 atoms and $\epsilon=0.5$ and $0.75$ respectively}
\label{fig2}
\end{figure}

Now we consider the case of a magnetic impurity inside the corral and
analyze the correlation between the impurity and conduction electron spins.
The canonical RKKY calculation is perturbative in the exchange interaction 
$J$. In the present case there is no singular scattering as in the Kondo
model ($J<0$) and perturbation theory gives the correct result. An alternative way
of doing the calculation is to solve exactly a simpler interaction of the
Ising form $S_{imp}^{z}\psi ^{\dagger }({\bf R}_{imp})\sigma ^{z}\psi ({\bf R%
}_{imp})$. We have shown that in small systems the spin-spin correlations
along the $z$-direction evaluated with the Ising interaction reproduce
quantitatively the exact correlations calculated with the full isotropic
interaction\cite{HCB}. For the Ising-type interaction, the electron
propagator $G_{\sigma }({\bf r},{\bf r}^{\prime },\omega )$ in the presence
of the impurity can be evaluated by including and extra spin-dependent
scattering center in the matrix ${\bf g}$ of Eq. (\ref{g0}) or can be
written in terms of $G_{\sigma }^{0}({\bf r},{\bf r}^{\prime },\omega )$ as:

\begin{equation}
G_{\sigma }({\bf r},{\bf r}^{\prime },\omega ) =G_{\sigma }^{0}({\bf r},
{\bf r}^{\prime },\omega )+  
G_{\sigma }^{0}({\bf r},{\bf R}_{imp},\omega )\Sigma _{imp}(\omega
)G_{\sigma }^{0}({\bf R}_{imp},{\bf r}^{\prime },\omega )  \label{gimp}
\end{equation}
where

\begin{equation}
\Sigma _{imp}(\omega )=\frac{JS_{imp}^{z}\sigma }{1-JS_{imp}^{z}\sigma
G_{\sigma }^{0}({\bf R}_{imp},{\bf R}_{imp},\omega )}  \label{sig}
\end{equation}

This expression is more appropriate to interpret the results. The spin-spin
correlation defined as $C^{z}({\bf R}_{imp},{\bf r})=<S_{imp}^{z}\psi
^{\dagger }({\bf r})\sigma ^{z}\psi ({\bf r})>$ can be put into the form

\begin{equation}
C^{z}({\bf R}_{imp},{\bf r})=S_{imp}^{z}\int_{-\infty }^{\varepsilon
_{F}}d\omega [\rho _{\uparrow }({\bf r},\omega )-\rho _{\downarrow }({\bf r}%
,\omega )]  \label{comega}
\end{equation}
where the spin dependent local densities of states are $\rho _{\sigma }({\bf %
r},\omega )=-\frac{1}{\pi }%
\mathop{\rm Im}%
G_{\sigma }({\bf r},{\bf r},\omega +i0^{+})$. If a second impurity is placed
at coordinate ${\bf R}_{imp}^{\prime }$ the effective impurity-impurity
coupling is proportional to $C^{z}({\bf R}_{imp},{\bf R}_{imp}^{\prime })$.

\begin{figure}
\onefigure{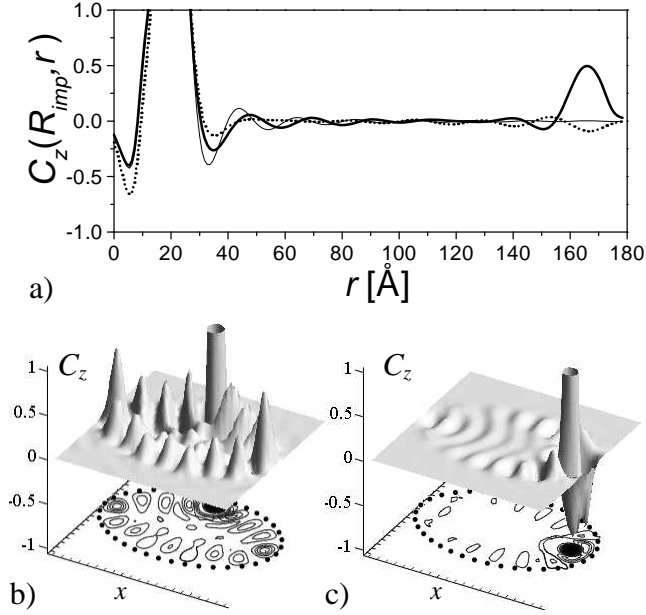}
\caption{a) Spin correlation function $C_z$ along the principal axis with
the
impurity at the left focus for the {\it at resonance} (thick
line) and {\it off resonance} (dotted line) situations as compared to
the open surface correlations (thin line). b) {\it At resonance} $C_z$ with 
the impurity far from the foci; c) {\it off resonance} $C_z$ with the
impurity at the focus} 
\label{fig3}
\end{figure}

Figures~\ref{fig1} and \ref{fig3} illustrate the behavior of the correlation
function. In
Fig. (3a) a cut of $C^{z}({\bf R}_{imp},{\bf r})$ along the principal axis
of the elliptical corral with the impurity at one focus is shown and
compared with the same quantity in the absence of the corral, in fig (3b)
the correlation function for the impurity away from the foci is presented.
In this last situation again, two well pronounced maxima close to the foci
are obtained. At very short distances $C^{z}({\bf R}_{imp},{\bf r})$ depends
on the high energy cutoff, however as $|{\bf R}_{imp}-{\bf r}|$ increases it
rapidly converges to a stable value.

Comparing the results of $C^{z}({\bf R}_{imp},{\bf r})$ with the local
density of states at the Fermi energy (fig (2c)), it is clear that the
correlation function reproduces the structure of the unperturbed density of
states $\rho ^{0}({\bf r},\varepsilon _{F})$. To understand this rather
intuitive result, we consider a simple limit in which both, the coupling constant $J$ and the
resonance widths $\gamma$ inside the corral are small compared to the energy separation
$\Delta \varepsilon$ between resonances. Note that for all states around a resonance, the
wave function inside the corral is essentially of the same form, differing only outside
the corral. To lowest order, the impurity shifts the position of the resonances by
$\eta _{\nu ,\sigma}\simeq JS_{imp}^{z}\sigma |\varphi _{\nu }({\bf R}_{imp})|^{2}$.
Inserting
this result into Eqs. (\ref{gimp}) and (\ref{comega}), after some algebra
and assuming that the Fermi energy coincides with one of the resonant states ({\it at
resonance} situation) we simply obtain $C^{z}({\bf R}_{imp},%
{\bf r})=A |\varphi _{\nu _{F}}({\bf r})|^{2}$ where $A$ is a proportionality constant.
If the Fermi energy lies between two resonant states ({\it off resonance} situation), to
lowest order the spin-spin correlation is zero. The high response of the conduction electrons
obtained for $r\to R_{imp}$ is due to higher order corrections.
The occurrence of magnetic mirages of impurities depends on whether the Fermi
energy lies at, or close to, one resonance or between two of them, an effect also observed in
spectroscopic mirages of Kondo impurities\cite{Mano}. In
quantum corrals, since the Fermi energy is fixed by the bulk material and
the position of the resonances depends on the corral size, the system can be
designed to have the Fermi level in any of the two situations. To show the
behavior of the conduction electrons response when the Fermi energy is
off resonance, instead of changing the corral size, we have shifted the
Fermi energy to 428 meV which corresponds to a valley between two resonances
in the local density of states. In this case the spin-spin correlation $%
C^{z}({\bf R}_{imp},{\bf r})$ shown in figs. (3a) and (3c) presents small amplitude
oscillations and an antiferromagnetic correlation with the conduction electrons at
the empty focus\cite{HCB}.

As in the case of the spectroscopic mirage of Kondo impurities\cite{agam,Egler,BW},
the magnetic mirage obtained in elliptical corrals is
due to a symmetry of the system that generates a local density of states at
the Fermi level with two pronounced maxima close to the foci. This structure
enhances the impurity-impurity interaction if the impurities are placed
precisely at, or close to, these points. In all nanoscopic systems however,
the local density of states at $\varepsilon _{F}$ shows some structure that
will tend to reinforce the magnetic interactions between impurities even if
the system geometry is not so special provided that the amplitude 
$|\varphi _{\nu }({\bf R}_{imp})|^{2}$ of the states close to the Fermi level is not too
small. In figs. (4a) and (4b) we present the conduction
electron response for circular and square corrals. A comparison of these
results with that of the impurity at an open surface (fig (1a)), clearly
shows the enhancement of the amplitude of the oscillations in the spin- spin
correlation due to confinement effects. Similar effects are expected in
small metallic grains. During the last decades a lot of work has been done
in small metallic clusters with sizes varying from that of a few atoms to
macroscopic grains. Magnetic impurities in these systems should
behave as in the examples presented above.

\begin{figure}
\onefigure{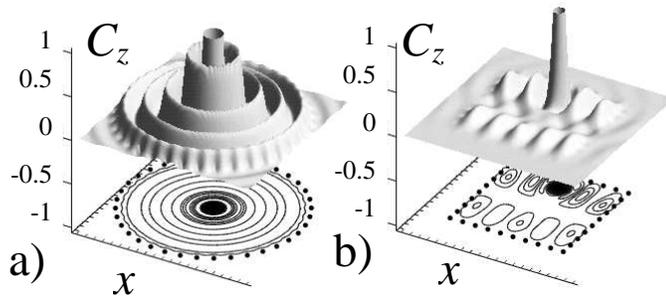}
\caption{Spin correlations $C_z$ for a circular corral with the
impurity at the center (a) and for a square corral (b). }
\label{fig4}
\end{figure}

\begin{figure}
\onefigure{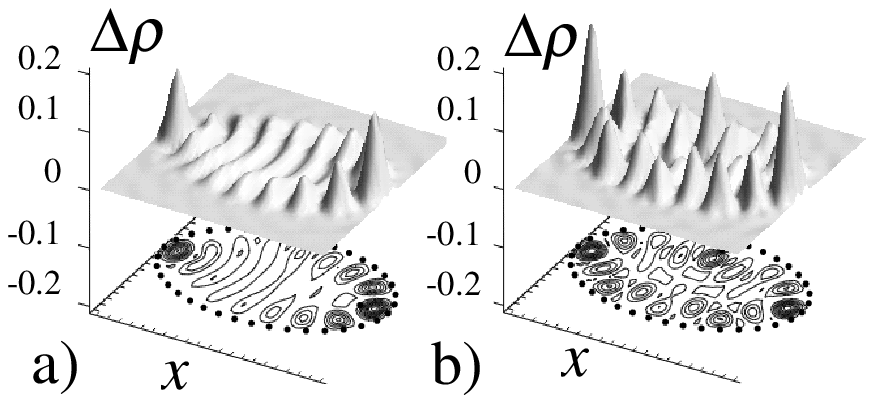}
\caption{$\Delta \rho$ for the impurity at the focus (a) and
far from it (b), for the {\it at resonance} situation}
\label{fig5}
\end{figure}

Recent advances in STM techniques that allow for the injection of
spin-polarized electrons \cite{spstm} could be used for a direct measurement
of the effect in quantum corrals. The spin-polarized current is obtained by
using ferromagnetic Fe or Co tips. A small magnetic field ${\bf B}$ at low
temperatures could be used to orient the impurity spin parallel or
antiparallel to the tip magnetization and the difference in the current
could be measured at each point. If the impurity spin is parallel to the
magnetization tip, the total current is proportional to $a\rho _{\uparrow }(%
{\bf r},\varepsilon _{F})+b\rho _{\downarrow }({\bf r},\varepsilon _{F})$
where $a$ and $b$ are characteristic of the tip and are proportional to its
majority and minority spin densities of states. For the reverse orientation
of the impurity spin the current is $b\rho _{\uparrow }({\bf r},\varepsilon
_{F})+a\rho _{\downarrow }({\bf r},\varepsilon _{F})$ and the difference is $%
\Delta j\propto (a-b)[\rho _{\uparrow }({\bf r},\varepsilon _{F})-\rho
_{\downarrow }({\bf r},\varepsilon _{F})]$. Although the last bracket is not
the spin-spin correlation it has the same structure. In fig.~\ref{fig5}
the function $\Delta \rho =[\rho _{\uparrow }({\bf r},\varepsilon _{F})-\rho
_{\downarrow }({\bf r},\varepsilon _{F})]$ is shown for two different cases.

In summary, we have studied the response of conduction electrons with
quasi-confined (resonant) states to a local perturbation produced by a
magnetic impurity. We have analyzed, using realistic parameters, the case of
impurities in quantum corrals. The main conclusions are:

{\it i)} If the Fermi energy is at resonance, or close to a resonant state,
the response of the conduction electrons to the impurity is strongly enhanced.

{\it ii)} If the exchange interaction is small, compared with the typical
energy difference of resonances, away from the impurity position the
spin-spin correlation function reproduces the structure of the local density
of states at the Fermi energy.

{\it iii)} Depending on the particular structure of the system, the
effective interaction between two impurities could be larger at large
distances than at short or intermediate distances. This occurs in elliptical
corrals where a magnetic mirage is formed due to the particular structure of
the local density of states at the Fermi energy.

{\it iv)} A direct measurement of the electron response to the impurity
could be made by injecting spin polarized electrons with a ferromagnetic STM
tip.

{\it v)} Similar effects, produced essentially by the confinement of
conduction electrons, are expected in other systems like small metallic
clusters or short carbon nanotubes.

\acknowledgments
We thank A.A. Aligia for providing the eigenstates of fig.~\ref{fig2}. This work
was partially supported by the CONICET and ANPCYT, grants N. 02151 and N.
06343. K. H. is fully supported by CONICET.

\end{document}